\documentstyle[11pt,newpasp,twoside,epsf]{article}
\def\edcomment#1{\iffalse\marginpar{\raggedright\sl#1\/}\else\relax\fi}
\reversemarginpar

\newcommand{\OIII}{\mbox{O\,{\sc iii}}}
\newcommand{\OI}{\mbox{O\,{\sc i}}}

\newcommand{\NII}{\mbox{N\,{\sc ii}}}
\newcommand{\SII}{\mbox{S\,{\sc ii}}}

\def\farcmin{\hbox{$.\!\!^{\prime}$}}

\begin{document}

\title{IRAS F02044+0957: radio source in interacting system of galaxies}

\author{O.V. Verkhodanov}
\affil{Special astrophysical observatory, Nizhnij Arkhyz, Russia}
\author{V.H. Chavushyan, R. M\'ujica and  J.R. Vald\'es}
\affil{Instituto Nacional de Astrof\'{\i}sica \'Optica y Electr\'onica, Puebla,
       M\'exico}
\author{S.A. Trushkin}
\affil{Special Astrophysical Observatory RAS, Nizhnij Arkhys, Russia}

\begin{abstract}
The steep spectrum of IRAS\,F02044+0957 was obtained with the RATAN-600 radio
telescope at four frequencies. Optical spectroscopy of the system components,
was carried out with the 2.1m telescope of the Guillermo Haro Observatory.
Observational data allow us to conclude that this object is a pair of
interacting
galaxies, a LINER and a HII galaxy, at $z=0.093$.

\end{abstract}


Trushkin \& Verkhodanov (1995) compiled  a list of about 750 objects
by using the CATS, as a result of cross-identifications of infrared
IRAS catalogues and the source catalogue of Texas survey at 365 MHz.
From the sample of steep spectra sources we selected those without
classification in  public databases (CATS, NED, ADS and LEDA). One
of these objects is IRAS~F02044+0957, identified with the NVSS radio
source J020706+101147.

Radio observations were carried out in April 23--25, 1999 with the
North Sector of the RATAN--600 telescope. The wide-band radiometer
complex at four frequencies: 2.3, 3.9, 7.7 and 11.2 GHz, was used.
The nearby non-variable radio source PKS 1345+12, was used as the
flux density calibration source. A linear fitting to the radio
spectrum was used to estimate a spectral index by the least square
method. Each point of the spectrum was weighted proportionally to
the value of $1/(\Delta\,S/S_\nu)^2$; where $\Delta\,S/S_\nu$ is
the relative error of the flux density. The radio spectrum is described
by a power-law $S_\nu[Jy] = 85\,\nu^{-0.94\pm 0.02}$~MHz.
The radio luminosity in the frequency interval from 365 to 10$^4$ MHz is
$L=9.381 \times 10^{34}$~\,erg~\,sec$^{-1}$~cm$^{-2}$ (assuming
$H_0$=64~\,km~\,Mpc$^{-1}$\,sec$^{-1}$ and$q_0$=0.8).

\begin{figure}
\plottwo{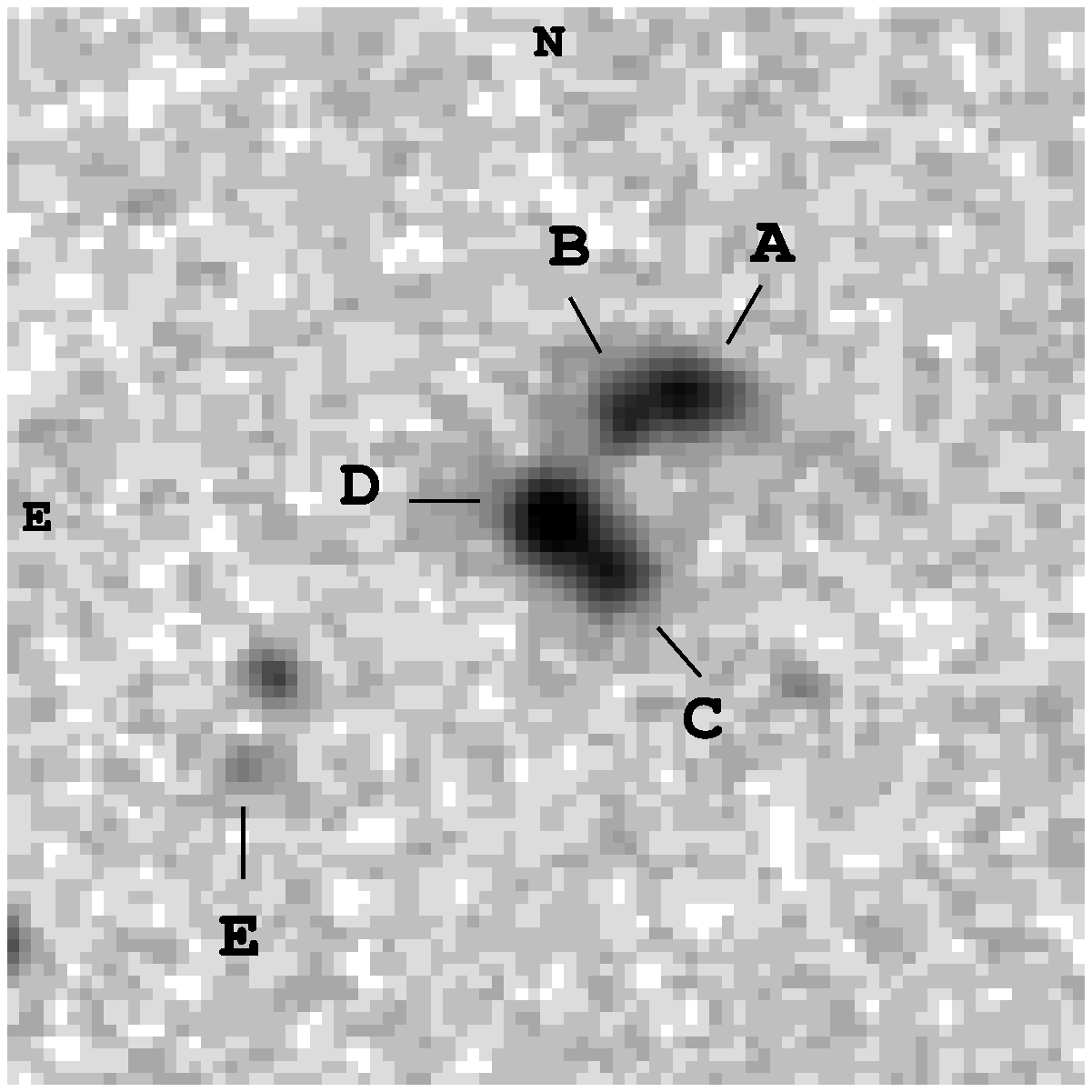}{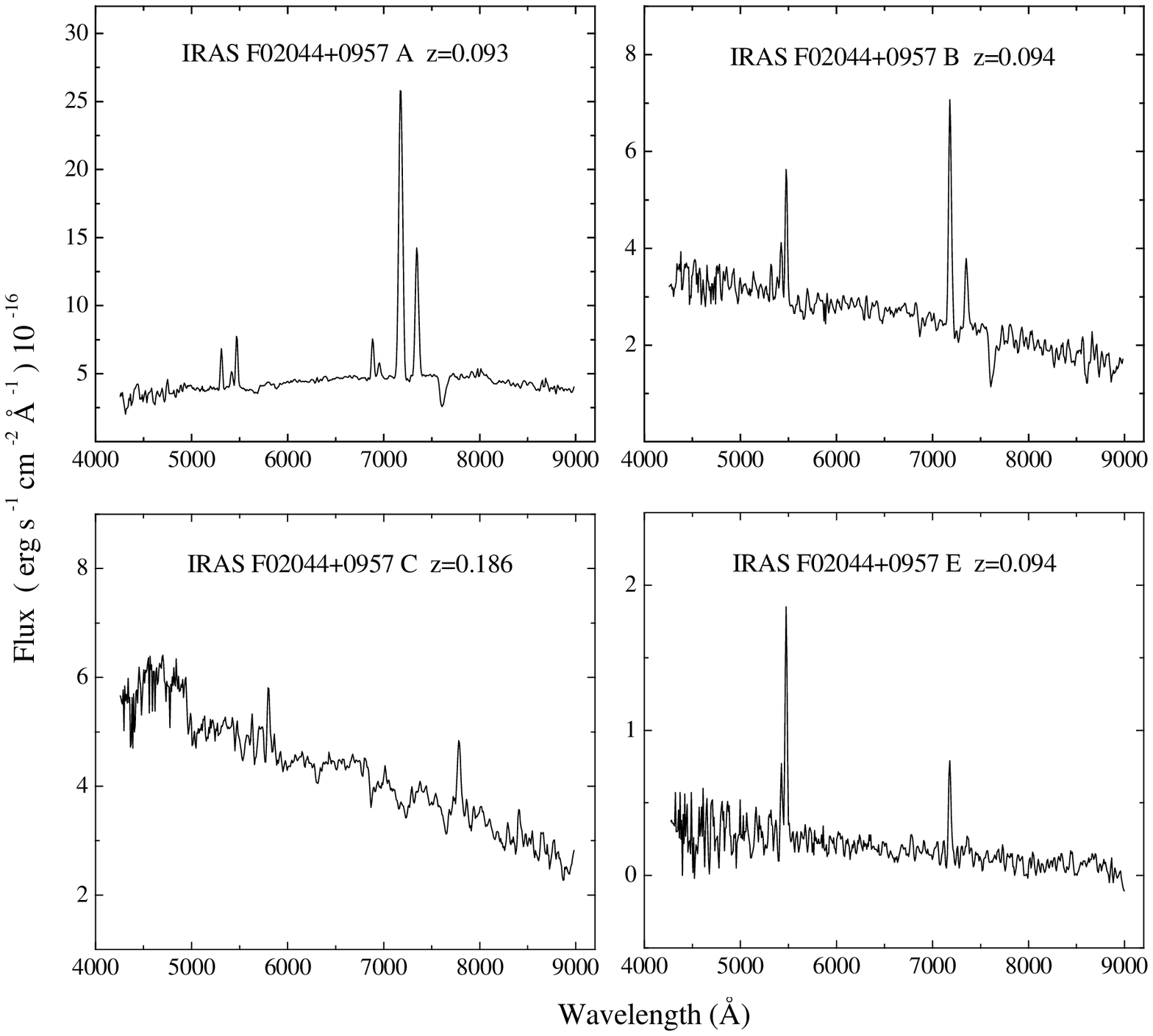}
\caption{DSS image of IRAS\,F02044+0957 and corresponding spectra}
\vspace*{-0.5cm}
\end{figure}

Optical spectroscopy of objects marked by A, B, C, D, and E in Fig. 1,
was obtained in August and November 1999. We used the 2.1m telescope of
the Guillermo Haro Observatory (GHO) in Cananea, Sonora, Mexico,
operated by the  National Institute of Astrophysics, Optics and
Electronics (INAOE). The Faint  Object Spectrograph and Camera (LFOSC)
(Zickgraf et al. 1997) was used. A setup, covering the spectral range
$4200-9000$\AA\, with a dispersion of 8.2\,\AA/pix was adopted. The
effective instrumental spectral resolution is about 15\,\AA.

The data reduction was done using the IRAF packages and included bias
and flat field corrections, cosmic rays cleaning, wavelength linearization
and flux  transformation.

The detailed spectroscopy showed that objects A and B are apparently
a pair of interacting emission-line galaxies at z=0.093. The second
pair (objects C and D) is a geometrical projection of a star (D) and
an emission line galaxy (C) at z=0.186. It is not possible that the
G type Main Sequence star, is the source of infrared and radio emission.

IRAS~F02044+0957 is located  at $16'$ from the center of the galaxy
cluster ZwCL 0203.6+1008 (Zwicky\,619).
The radius of the cluster is $12\farcmin6$. This means that
IRAS~F02044+0957 must be out the cluster boundaries.

In order to investigate the nature of the interacting system AB, we used
the diagnostic diagrams Log([\NII]$\lambda$6583/H$\alpha$) vs
Log([\OIII] $\lambda$5007/H$\beta$) and
Log([\SII]\-$\lambda\lambda$6717+6731/H$\alpha$)
vs Log([\OIII]$\lambda$5007/H$\beta$).

The values for component A, Log([\NII]\-/H$\alpha$)~=~$-$0.15 and
Log([\SII]/H$\alpha$)~=~$-$0.23, put the object on both diagrams in the AGN
region,  very close to the boundary with the HII region-like galaxies.
According to  the criteria for spectral classification proposed by Ho,
Filippenko \& Sargent (1997) (HFS97),  the component A is a LINER because
the value of [\OI]$\lambda$6300/H$\alpha$~=~0.26.
For component B, we obtained the values:
Log([\NII]\-/H$\alpha$)\,=\,$-$0.76 and
Log([\SII]/H$\alpha$)\,=\,$-$0.55. Therefore, according to HFS97,  the
component B is a HII galaxy. Summarizing, the interacting system AB is composed
by a LINER and a HII galaxy.

\acknowledgments

This work was partially supported by CONACyT \linebreak grants 28499-E,
J32178-E, and 32106-E.

\end{document}